\documentclass[10pt,halfline,a4paper]{ouparticle}
\usepackage{cite}
\usepackage{graphicx}

\begin{document}

\title{Deep learning systems as complex networks}

\author{\small 
\name{Alberto Testolin}
\address{Department of General Psychology, University of Padova\\
Via Venezia 12, 35131, Padova, Italy}
\email{alberto.testolin@unipd.it}
\and
\name{Michele Piccolini}
\address{Department of Physics and Astronomy, University of Padova\\
Via Marzolo 8, 35128, Padova, Italy}
\email{michele.piccolini@edu.unito.it}
\and
\name{Samir Suweis}
\address{Department of Physics and Astronomy, University of Padova\\
Via Marzolo 8, 35128, Padova, Italy}
\email{samir.suweis@unipd.it}}

\abstract{Thanks to the availability of large scale digital datasets and massive amounts of computational power, deep learning algorithms can learn representations of data by exploiting multiple levels of abstraction. These machine learning methods have greatly improved the state-of-the-art in many challenging cognitive tasks, such as visual object recognition, speech processing, natural language understanding and automatic translation. In particular, one class of deep learning models,  known as \emph{deep belief networks}, can discover intricate statistical structure in large data sets in a completely unsupervised fashion, by learning a generative model of the data using Hebbian-like learning mechanisms. Although these self-organizing systems can be conveniently formalized within the framework of statistical mechanics, their internal functioning remains opaque, because their emergent dynamics  cannot be solved analytically. In this article we propose to study deep belief networks using techniques commonly employed in the study of complex networks, in order to gain some insights into the structural and functional properties of the computational graph resulting from the learning process.}
\date{\today}
\maketitle
%

\section{Introduction}

Recent strides in artificial intelligence research have opened tremendous opportunities for technological development. In particular, the last decade has been marked by the so-called ``deep learning revolution'', which is having strong impact both for scientific investigation and for engineering applications \cite{LeCun2015}. Deep learning allows building artificial neural networks composed of many processing layers, which can learn high-level representations of the data by exploiting multiple levels of abstraction \cite{Hinton2007a}. To differ from conventional machine-learning techniques, this allows to automatically discover intricate statistical structure in large datasets without the need for domain expert knowledge: the relevant features needed to describe the data distribution are directly learned by the machine from the raw input (e.g., pixels values in a digital image). An intriguing aspect of deep learning systems is that they are inspired by neuronal networks in biological brains: information processing occurs in a parallel and distributed fashion \cite{Rumelhart1986a}, thereby allowing cognitive abilities to emerge from the orchestrated operation of many simple, non-linear processing units \cite{McClelland2010,Testolin2016a}.

Deep learning has dramatically improved the state-of-the-art in challenging cognitive tasks, such as image classification \cite{Krizhevsky2012,he2016deep}, speech recognition \cite{Mohamed2012a}, natural language understanding \cite{Collobert2011} and even high-level reasoning \cite{Mnih2015,Silver2016}. It is currently employed by all major IT companies (Google, Facebook, Microsoft, Apple, just to mention a few) to automatically extract knowledge from large digital datasets, and it is achieving impressive performance also in many other domains such as drug discovery \cite{Ma2015}, genomics \cite{Xiong2015}, high-energy physics \cite{baldi2014} and telecommunications \cite{Zorzi2015,testolin2014machine}.

However, despite the continuous progress and the widespread deployment in real-world applications of deep learning, there is still a limited comprehension of its working principles \cite{Baldassi2016}. How does these multilayer networks self-organize to solve a particular task? How is information represented in these systems? Is there a set of fundamental properties underlying the structure and dynamics of deep neural networks? 

Some insights into these challenging questions have been gained by inspecting deep learning systems with methods borrowed from neuroscience. For example, response profiles of individual neurons in deep networks often exhibit an impressive match with neurophysiological data \cite{Lee2008,Testolin2017,Testolin2017b,Zorzi2017a}. Similarly, at the neuronal population level it has been shown that the representational space developed by deep networks has a striking overlap with that observed in the inferior temporal cortex of the primate brain \cite{Guclu2015,Kriegeskorte2015}. However, these empirical analyses are somewhat limited in scope because they do not allow to systematically assess structural and functional properties of these complex systems.

We believe that a fresh perspective on these issues can be provided by studying deep learning using the analytical and numerical techniques developed by network science \cite{Albert2002,Newman2010}, which have already provided very useful in neuroscience research \cite{Bressler2010,Bullmore2009,Park2013,Medaglia2015}.
Indeed, in deep learning even knowing perfectly how a single neuron (node) of the network works does not allow to understand how learning occurs, why these systems work so efficiently in many different tasks, and how they avoid getting trapped in configurations that deteriorate computational performance \cite{Baldassi2016,lin2017,mhaskar2017}. In these models, interactions play a crucial role during the learning process, therefore a step forward toward a more comprehensive understanding of deep learning systems is their study also in terms of their emerging topological properties \cite{agliari2015topological}.
For example, a first characterization of deep networks can be done through several statistical graph properties: connectance, degree distribution, strength distribution, and  other centrality measures that are now standard in network theory \cite{Newman2010}. In particular, we expect that these properties encapsulate relevant information about the learning process of the system. Can we unveil some general relationship between the function (learning outcome) and the structure (topology) of the network? What does depend on the learning task, and what is instead independent of the specific distribution of the input data?

In summary, the aim of the present work is to show that a network science perspective on deep learning may enlighten some of these relationships. We also make available the source code of our software analyses in order to promote the use of the analytical methods we describe.

\section{Deep Neural Networks}
Strictly speaking, the main goal of deep learning is to support the creation of ``intelligent'' machines that can autonomously learn from experience. Indeed, according to the view promoted by neural network models, perceptual and cognitive phenomena can be conceived as the evolution over time of a complex system of interconnected units that self-organize according to physical principles \cite{McClelland2010}. Within this framework, the pattern seen in overt behavior (macroscopic dynamics of the system) emerges from the coordination of subcognitive processes (microscopic dynamics of the system), such as the propagation of activation and inhibition among simple, but non-linear, processing units.
The computational properties of neural networks have been investigated since the dawn of artificial intelligence research \cite{McCulloch1943,rosenblatt1958}, but only recently the computational power of these systems has been fully unleashed: the major achievement of deep learning algorithms has been to show that artificial neural networks can learn a hierarchy of increasingly more complex concepts, with each concept defined through its relation to simpler concepts (for an historical review, see \cite{Schmidhuber2015}).

In the following, we will first briefly review the basic theory of learning in deep neural networks, and we will then describe the details of the deep learning model used in our study. For a recent and comprehensive survey about the topic, the reader could refer to \cite{Goodfellow2016}.

\subsection{Theoretical Background}
Artificial neural networks can be formally described using the theory of probabilistic graphical models \cite{jordan2001,koller2009}, which provides a general framework to model the stochastic behavior of a large number of interacting variables. Learning in probabilistic graphical models can be framed within two different settings: In \emph{discriminative} models, the goal is to model conditional distributions over a set of output (target) variables, whose values are specified by explicitly labeling each pattern given as input to the system. This  approach is usually referred to as \emph{supervised learning}, because the system is always guided by an external teacher who provides the correct labels. Classification, discrimination and regression problems can be easily framed within this scenario, and are usually solved by applying feed-forward, convolutional deep neural networks trained with error backpropagation (e.g., \cite{Krizhevsky2012}).
In \emph{generative} models, instead, the aim is to capture the joint distribution of all the variables in the system, thereby including also the input variables. This learning modality is usually described as being \emph{unsupervised}, because there are no correct labels that must be associated with each input pattern. The goal is rather to build an internal model of the environment, that is, to discover a set of \emph{latent features} that compactly describe the statistical correlations observed between the variables at play. Clustering, density estimation and dimensionality reduction problems can be framed within this scenario.

In this article we will focus on the latter approach, because generative neural networks can be more naturally characterized using the physical formalism of statistical mechanics, as we will highlight in the following.

\subsubsection{Boltzmann Machines}
\begin{figure}[t]
\includegraphics[width=\textwidth]{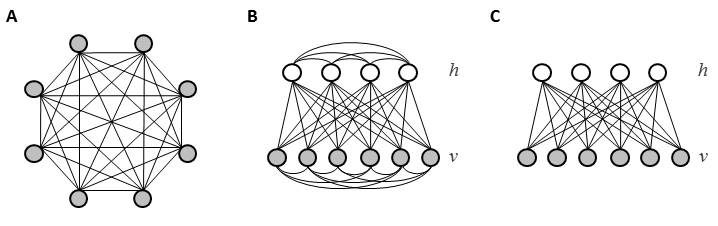}
\caption{Graphical representation of several recurrent neural network architectures. (A) Hopfield network, which is a fully-connected graph. (B) Boltzmann machine, which is a fully-connected graph with two separate sets of units: the visible neurons \emph{v} are used to perceive input patterns, while the hidden neurons \emph{h} capture the statistical structure observed in the data. (C) Restricted Boltzmann machine, where intra-layer connections are removed in order to form a bipartite graph.}
\label{network_architectures}
\end{figure}

Generative models can be implemented using different types of probabilistic graphical models. One of them is the Boltzmann machine \cite{ackley1985}, which is an undirected model (i.e., edges are symmetric, implying a bidirectional flow of information between the nodes) that has been traditionally defined using concepts borrowed from statistical physics. In particular, following the seminal model introduced by Hopfield \cite{Hopfield1982}, it can be shown that this type of fully-connected, recurrent networks (see Fig. \ref{network_architectures}A) can develop a point-attractor dynamics, which can be analyzed using techniques inspired by the study of pattern formation in physical systems composed by many interacting units. This allows to draw a useful analogy between physical systems with a metastable behavior and information processing systems that implement content-addressable associative memories: each local energy minima in a metastable physical system can be interpreted as an embodiment of a ``prototype'' in an associative memory, where the aim is to store as much information as possible in the form of static configurations of a set of variables. If each configuration is defined by the actual state of the variables, it is possible to recall previously stored information by giving as input to the network a partially observed state (which would correspond to a specific initial condition acting as content-specific ``search key'') and letting the system settle into a stable state, thereby completing the missing values of the remaining variables according to the closest prototype (that is, by converging to the closest attractor).

Modeling phase transitions in such physical systems has been a longstanding issue in statistical mechanics, which has been explored especially in the context of Ising models. An Ising model is a collection of $K$ binary variables ($\sigma_i = {+1,-1}$ representing the spins of the atoms in the up (1) or down (-1) state in a ferromagnetic material) arranged into a two-dimensional lattice, which are magnetically coupled to each other. This can be mathematically represented by assigning an \emph{energy function} to the state of the whole lattice $\sigma={\sigma_1,...,\sigma_K}$ $\in\mathbb{R}^K$, given the coupling $J_{mn} = J$ if m and n are neighbors and $J_{mn} = 0$ otherwise, and the external magnetic field $H$:
\begin{equation}
	E(\sigma; J,H) = - \frac{1}{2} \sum\limits_{m,n} J_{mn}\sigma_m \sigma_n - \sum\limits_n H \sigma_n 
\label{eqn:IsingEnergy}
\end{equation}
At equilibrium, spins try to align with the external field and to get parallel each other in order to fulfill the minimum energy principle, i.e. minimize $E(J,H)$. The stationary state distribution for the spin system at temperature $T$ is given by the usual Boltzmann distribution:
\begin{equation}
	P({\sigma}| \beta, J, H) = \frac{1}{Z(\beta, J, H)} e^{-\beta E(\sigma;J,H)}
\label{eqn:Boltzmann}    
\end{equation}
where $\beta =  \frac{1} {k_B T}$ defines the inverse temperature of the system ($k_B$ can be set to $1$), and $Z(\beta, J, H) = \sum\limits_{\sigma} e^{-\beta E(\sigma; J,H)}$ is the \emph{partition function} that assures the normalization of the distribution.

A Boltzmann machine is a generalization of the Ising model, where all units are connected to each other by bi-directional links, i.e. in this case the couplings are given by a fully connected matrix $W$. If we now call $x$ the state of the machine, where each unit $i$ can be off ($x_i=0$) or on ($x_i=1$), then we can write the energy gap of the $j$-th unit, defined as the difference between the energy of the whole system with the $j$-th ``off'' and its energy with $j$-th ``on'' by: 
\begin{equation}
	\Delta E_j =\sum\limits_{i} W_{ij}x_i
\label{eqn:BoltzmannEnergy}
\end{equation}
where $W$ is the matrix of synaptic connection weights, which are symmetric (i.e., $w_{ji}=w_{ij}$) and which define the reciprocal interactions between all neurons in the network. If $\Delta E_j<0$ then the switch off of the $j$-th element decreases the total energy $ E(x;W,H)$. Therefore we can minimize the energy by evolving the system over time through stochastic dynamics, where each neuron $j$ changes its local state regardless of its previous state to:
  \begin{equation}
  x_i = \left\{
  \begin{array}{ll}
  1  & \text{with probability} \: \frac{1}{1+e^{-\beta\Delta E_j}} \\
  0 & \text{otherwise}
  \end{array}
  \label{eqn:activation}
  \right.
  \end{equation}
where the activation energy $\Delta E_j$ depends on the overall activation received by unit $j$ from its neighbors. Iteratively updating the state of each unit using this rule, the global system configuration is driven toward thermal equilibrium, that is, toward a state where the energy is locally minimized, thereby following the Boltzmann distribution $P({x} \vert W)$ with energy $E(x;W)$ (as in Eq. (\ref{eqn:Boltzmann})). In order to avoid the system being trapped in local minima with relatively high-energy, the overall temperature of the system can be gradually decreased, thereby mimicking the annealing process in physical systems \cite{kirkpatrick1983}.

Perhaps the most interesting property of Boltzmann machines is their ability to ``learn'' by modifying the connections strength between units in response to the statistical properties of an external signal that is provided as input to the system. To this aim, the whole network can be partitioned into two distinct functional subsets of units (see Fig. \ref{network_architectures}B): a set of \emph{visible} neurons $v$, which are the interface between the network and the external environment and therefore receive the input pattern (the ``observed'' data, in the language of graphical models), and a set of \emph{hidden} neurons $h$, which constitute the internal state of the network and which are used to capture useful ``features'' (i.e., statistical correlations) in the input variables.

Intuitively, the objective of learning is to discover a useful set of features, which would serve as latent variables that compactly encode the statistical structure contained in the input data. To this aim, during the learning phase all the network connections are initially set to small, random values, and then are gradually adjusted as the network observes new input data (the ``training examples''). The network modifies the strengths of its connections so as to construct an internal generative model that produces examples with the same probability distribution as the examples it is shown. At each learning iteration, all the visible units are clamped into a specific state provided by one training vector, and the hidden units activates according to Eq. \ref{eqn:activation}. This is called the \emph{positive} (or ``wake'') phase, because the system is driven by the input data \cite{hinton1995}. The visible units are then unclamped, and the network is left free to generate its own visible states by starting from a random state in the hidden units activations. This is called the \emph{negative} (or ``sleep'') phase, because the system is driven by its own internal model and tries to generate plausible configurations in the visible layer. Following these two phases, the model parameters (i.e., the connection weights) are updated by maximizing the agreement between the empirical correlations among visible and hidden units resulting from the positive phase and those resulting from the negative phase \cite{ackley1985}.

Formally, given a set of training examples clamped to the visible neurons $\{{v}^{(n)}\}_1^N$ we would like to adjust the connection weights $W$ such that the samples generated by the network are well matched by those provided in the training distribution. To this aim, we can define learning as a maximum likelihood problem, where a set of model parameters $W$ has to be adjusted in order to maximize the likelihood of the sampled data. By performing gradient descent on the empirical negative log-likelihood of the training data, we can analytically derive the equation for updating the connection weights (model parameters) at each learning iteration. Crucially, the derivative of the log-likelihood of a training example with respect to the weight $w_{ij}$ turns out to be surprisingly simple:
  \begin{equation}
  \frac{\partial \log P({v,h}; \beta, J, H)}{\partial w_{ij}} = \langle v_i h_j \rangle_{data}  - \langle v_i h_j \rangle_{model}
  \end{equation}
where the angle brackets are used to denote expectations under the distribution specified by the subscript that follows, that is,  under the empirical data distribution (positive phase) and under the model distribution (negative phase). This leads to a very simple learning rule for updating each connection weight of the network:
    \begin{equation}
  \Delta w_{ij} = \epsilon(\langle v_i h_j \rangle_{data}  - \langle v_i h_j \rangle_{model})
  \end{equation}
where $\epsilon$ is a small constant representing the learning rate.

\subsubsection{Restricted Boltzmann Machines}
Although interesting from a theoretical perspective, Boltzmann machines are seldom used in practice due to their extremely high computational complexity. Indeed, these models have an intractable partition function, which prevents the exact computation of the likelihood gradient. This issue can be mitigated using mean-field approximations \cite{welling2002}, but computing the model's expectations $\langle v_i h_j \rangle_{model}$ still remains computationally demanding.

A more effective approach has been instead to constrain the connectivity of the network, moving from a fully-connected topology to a bipartite graph (see Fig. \ref{network_architectures}C). By removing all the intra-layer connections we obtain a Restricted version of a Boltzmann Machine (RBM), where all the neurons in the same layer are conditionally independent given the state of neurons in the opposite layer. This allows to enormously speed-up learning, for example by exploiting efficient implementations of Monte Carlo methods based on parallel Gibbs sampling\cite{hinton2002}. For example, when the neurons in one layer are clamped to a particular state (e.g., the visible neurons $v$ are clamped to one training example), the activation probability of all the neurons in the other layer can be efficiently computed in one parallel step: 
  \begin{equation}
  P(h | v) = \prod_{i}{P(h_i | v)}
  \end{equation}
where:
\begin{equation}
  P(h_i = 1 | v) = \frac{1}{1+e^{-\sum\limits_{j} w_{ji} v_j }}
  \end{equation}
and $P({v,h} \vert 1, J, H)=P(h | v)P(v)$, omitting for simplicity the dependence on the parameters.
\subsubsection{Deep Belief Networks}
\begin{figure}[tb]
\centering
\includegraphics[width=0.4\textwidth]{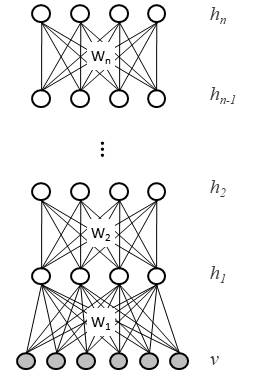}
\caption{Graphical representation of a deep belief network built as a stack of restricted Boltzmann machines.}
\label{dbn_architecture}
\end{figure}

A groundbreaking discovery is that RBMs can be used as building blocks to build more complex neural network architectures, where the hidden variables of the generative model are organized into layers of a hierarchy (see Fig. \ref{dbn_architecture}). These models are usually referred to as Deep Belief Networks (DBNs) \cite{hinton2006fast,salakhutdinov2015learning}.
Such systems are built by stacking together multiple RBMs, which are learned in a layer-wise fashion, that is, the $h_n$ layer is trained after training is completed for the $h_{n - 1}$ layer. In this way, the hierarchical generative model is built at separate stages, first starting with simpler features that are kept fixed in order to subsequently learn the more complex ones. After the first RBM has been learned, the activities of its hidden neurons are used as input for a second RBM, with the aim of extracting higher-order correlations from the original data. The main intuition behind these powerful architectures is that, by training a generative layer using as input the hidden causes discovered at the previous layer, the network will progressively build more structured and abstract representations of the input data. Importantly, architectures with multiple processing levels efficiently encode information by exploiting re-use of features among different layers: simple features extracted at lower levels can be successively combined to create more complex features, which will eventually unravel the main causal factors underlying the data distribution \cite{zorzi2013,Hinton2007a}. Indeed, it has been shown that functions that can be compactly represented by a depth $k$ architecture might require an exponential number of computational elements to be represented by a depth $k-1$ architecture \cite{Goodfellow2016}. Moreover, adding a new layer to the architecture increases a lower bound on the log-likelihood of the generative model \cite{hinton2006fast}, thus improving the overall representational capacity of the network.

Thanks to its efficiency, the algorithm proposed by \cite{hinton2006fast} solves the problem of learning in densely connected networks that have many hidden layers. Moreover, when implemented on multi-core hardware (e.g., graphical processing units, GPUs) deep learning is practical even with billions of connections, thereby allowing the development of very-large-scale simulations \cite{raina2009}.

\subsection{Network architecture and learning details}
In our analyses, we considered a popular deep network architecture that has been shown to achieve excellent performance on the MNIST benchmark task of handwritten digit recognition \cite{hinton2006fast,testolin2013}. To assess the robustness of our methods, we also analyzed a single-layer RBM trained on a very different type of stimuli, that is, natural image patches \cite{Testolin2017b}.

\subsubsection{Training datasets}
For the MNIST, learning was performed over a large dataset containing 60000 images of handwritten digits \cite{lecun1998}. Each training example contains a digit between 0 and 9, represented as a gray-scale image of size 28x28 pixels. The input to the network was therefore a vectorized matrix of 784 real-valued elements in the range [0,1].
For the natural image patches, learning was performed over a large dataset containing 80000 images of natural scenes \cite{snavely2006}. Each training example contains a small portion of a natural landscape, represented as a gray-scale image of size 40x40 pixels. The input to the network was therefore a vectorized matrix of 1600 real-valued elements in the range [0,1].

\subsubsection{Learning parameters}
For the MNIST, the DBN was composed of a stack of three RBMs. The visible layer contained 784 neurons, each corresponding to one pixel in the input image. Hidden layers had, respectively, 500, 500 and 2000 neurons, for a total of about 1.6 million connections in the whole DBN.
Learning was performed using one-step contrastive divergence, with a fixed learning rate $\epsilon=0.1$, a momentum coefficient of 0.9 and a weight decay factor of 0.0004. The latter two hyperparameters serve as regularizers for the loss function in order to minimize the risk of overfitting the training data: for additional details, the reader is referred to \cite{hinton2012}.
For the natural images, the RBM had 1600 neurons in the visible layer and 1000 neurons in the hidden layer. Learning hyperparameters were the same adopted for the MNIST model.

Both models were implemented using MATLAB by exploiting an efficient code tailored for graphic processors \cite{testolin2013}\footnote{The complete source code can be found at http://ccnl.psy.unipd.it/research/deeplearning}.

\subsubsection{Training Time}
Each layer of the DBN was trained for 50 epochs, where each epoch corresponds to a full sweep over all patterns in the training set. We verified learning convergence by monitoring the average reconstruction error after each epoch \cite{hinton2012}. After 50 epochs, the reconstruction error did not significantly further improve and most of the neurons already developed structured receptive fields.

\section{Network-based Analyses}
In deep learning systems the initial processing architecture is fairly generic. For example, in our DBN with $k$-1 hidden layers, it corresponds to a fully-connected (connectivity $C=1$) $k$-partite \cite{gary2008} graph with random weights drawn from a Gaussian distribution of mean zero and standard deviation $\sigma$, i.e. $w_{ij}\sim\mathcal{N}(0,\sigma)$. As a result of learning, complex structural patterns gradually emerge, that \emph{a priori} depend also on the input given for the training algorithm that we have described in detail before.

In the case of input digit images, we are dealing with a multilayer neural network composed by 3 stacked RBMs. The analysis of the network properties can be performed both on the DBN seen as one aggregate graph, and on each of the three bipartite networks between pairs of layers.
By denoting with $W_{a}$ ($a=1,2,3$) the weighted bipartite network between layers $v-h1,h1-h2,h2-h3$, respectively (see Figure \ref{dbn_architecture}), we can write the adjacency matrix $W$ of the whole DBN as:
\begin{equation}\label{W}
W=\left( \begin{array}{cccc}
\mathbb{O} & W_{1} & \mathbb{O}  & \mathbb{O} \\ 
W_{1}^{Tr} & \mathbb{O} & W_2 & \mathbb{O} \\ 
\mathbb{O}  & W_{2}^{Tr} & \mathbb{O} & W_{3}\\ 
\mathbb{O}  & \mathbb{O} & W_{3}^{Tr} & \mathbb{O} \end{array} \right)
\end{equation}
From $W$ we then calculate topological properties of both the intra-layers and the whole network. In particular, we are interested in studying the structure of subnetworks composed by groups of nodes that have characteristic functional properties (given by their receptive fields, see next section). In this way we try to infer topological signatures of the functional roles of the nodes. We first calculated the distribution of the network degrees, strengths and weights emerging as a result of the learning process in each layer $h_{i}$, with $i=1,2,3$. We also calculated the overlap \cite{Newman2010} between each pair of layers, to see if group of nodes are less or not activated. We finally computed the average degree, strength, coefficient of variation and average nearest neighbors of the subnetworks formed by nodes with similar functional properties. In all these cases we used a simplification threshold $\theta$ and set to zero all edge weights smaller than $\theta$, in order to obtain a non-weighted graph to work on.

\subsection{Neuronal Receptive Fields}
It is useful also to introduce the notion of \textit{receptive field}\cite{zorzi2013}, which allows to have a straightforward visualization of how a given neuron ``sees'' the input, i.e. it incorporates the functional role of that neuron given the input. In other words, the receptive field of a neuron represents the type of visual feature that has been extracted during learning. Since neurons in the visual layer are directly clamped to the input pattern, we can define the concept of receptive field only for neurons that live in upper, hidden layers.

\begin{figure}[t]
 \centering
 \includegraphics[width=0.3\linewidth]{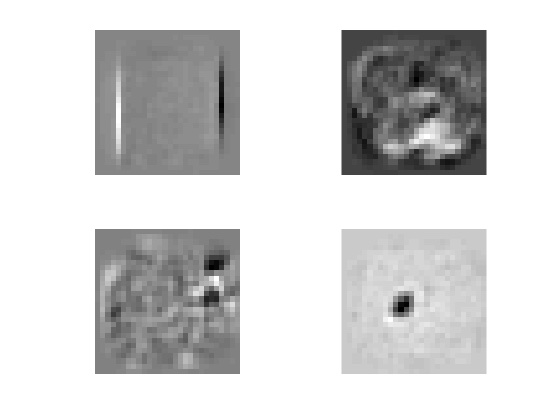}
 \caption{Example of four different receptive fields emerging in the hidden neurons.} \label{rfexample}
\end{figure}

The receptive field for a neuron of the first hidden layer is just the visual representation of the weights of its links toward the neurons in the input layer below. To plot the receptive field of the $j$-th neuron in the layer $h1$ we just need $W_1$, and extract the weights of the links that start from node $j$ in $h1$ and arrive to all nodes in the layer below ($v$), i.e. the vector $w^1_{i,j}$ for $i=1,...,n_1$ where $n_1$ is the number of nodes in the $v$-layer while $j$ is fixed. We can then reshape this vector to the original input square matrix form (whose dimension is $28 \times 28$ pixels for the case of the MINST handwritten digits). Each $i$-th pixel value is represented in a gray-scale color from the minimum (black) value to the maximum (white) value (receptive field samples are shown in Figure \ref{rfexample}).
These receptive fields are informative about the neuron's function in the sense that they suggest which features of the input images mostly activate the neuron. For example, a neuron could be particularly sensitive to straight, vertical lines in the peripheral side of the image, while being completely indifferent to analogous straight lines drawn at the bottom of the image (e.g. the first of the 4 images in Figure \ref{rfexample}). Other neurons could be excited by circular-ish ring shapes, while being anti-correlated with the space inside this ellipse (see 4th receptive field). Some neurons do not encode localized features, as they assume a distribution of weights that makes the receptive field a fuzzy blurred blob (as in the cases of the 2nd and 3rd images in the figure).

For neurons that lie in the $h2$ and $h3$ layers, we define the receptive field in an indirect way, since there is no direct linkage between the neuron and the visible layer. For a hidden layer 2 neuron (say, $l$), we first make a weighted average of the receptive fields of the neurons in the hidden layer 1, using the weights $w^{2}_{j,l}$ leaving from the $l$-th neuron and linking it with the $j$ neurons below. Then, we plot the 2-dimensional receptive field of the neuron as $\sum\limits_{j} w^{1}_{i,j} w^{2}_{j,l}$ ($l$ is fixed, while $i$ varies over the nodes in the visible layer), obtaining an image of what pixels our $l$-th neuron in the second hidden layer is responsible for. For a neuron $m$ in the third layer we proceed in the same way, by plotting: $\sum\limits_{j,l} w^{1}_{i,j} w^{2}_{j,l} w^{3}_{l, m}$. Results are somehow similar to the receptive fields observed in the first hidden layer, but we can see that at deepest layers the receptive fields become more ``structured'' and complex.

\subsection{Clustering of Receptive Fields}
In order to automatically group neurons with similar functional properties, we implemented a hierarchical clustering algorithm based on the earth mover's distance \cite{rubner2000earth}.
This distance metric defines the distance between two distributions as the minimal cost that must be paid to transform one distribution into the other, and it is widely used for content-based image retrieval. We adopted this metric because it is based on a solution to a transportation problem from linear optimization, for which efficient algorithms are available, and it guarantees a reasonable precision in measuring the visual similarity of two gray-scale images. Intuitively, the earth mover's distance is computed as follows: every pixel is represented by a certain number of ``pebbles'', which is an integer number corresponding to the the gray level of that pixel. After normalizing the two images to have the same number of ``pebbles'', the distance between them is computed as the minimum cost of matching the pebbles between the two images, which can be formalized and solved as a transportation problem \cite{rubner2000earth}. After computing the distance matrix between all receptive fields, hierarchical clustering was performed by creating a tree structure based on the Euclidean distances between all rows in the matrix. A dendrogram was finally produced by first computing the optimal ordering of the tree leaves using the \texttt{optimalleaforder} MATLAB function, and then calling the \texttt{dendrogram} function by setting to 20 the maximum number of leaf nodes. This resulted in a manageable number of receptive field clusters, at the same time limiting the creation of singletons or clusters with only few elements.

\section{Results}

\subsection{Weights Distribution}
\begin{figure}[t] 
 \centering
 \includegraphics[width=1\linewidth]{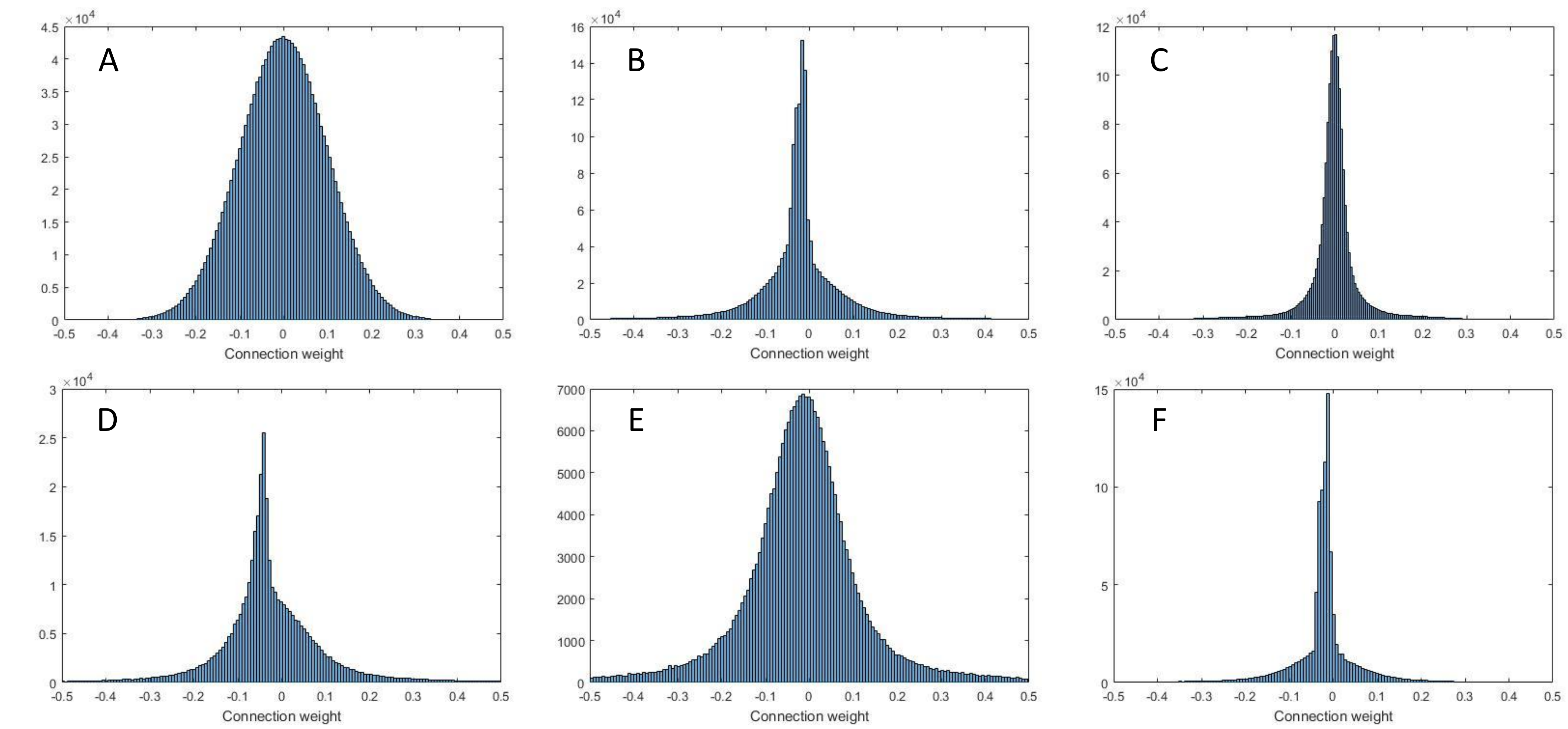}
 \caption{Distribution of the edge's weights for: A) the whole network before learning (initialization); B) the whole network after learning on the MNIST dataset; C) the network after learning on the natural images dataset; D) the first layer $v-h1$, E) second layer $h1-h2$ and F) third layer $h2-h3$ bipartite networks after learning on the MNIST dataset.} \label{WeightsALL}
\end{figure}

By analyzing the distribution of the edge weights $W$ of the whole deep network before and after learning, we observed a clear increase of inhibitory (negative) interactions, highlighted by a shift of the weights mean toward negative values. Moreover, after learning the weights distribution is no longer Gaussian (compare panels A and B in Fig. \ref{WeightsALL}), due to the increase of the skewness of the distribution. This effect is mainly due to the change of edge weights in the first $v-h1$ and third $h2-h3$ bipartite networks (see panels D and F in Fig. \ref{WeightsALL}). Interestingly, the distribution of the weights in the second layer $h1-h2$ is still quasi-normal, and with an average close to zero (but still negative). In this case, the departure from a Gaussian distribution is mostly highlighted by an increase of the variance, which led to the emergence of long tails (see panel E in Fig. \ref{WeightsALL}). A similar result holds for the network trained on natural images. In this case, however, we do not observe a clear shift of the weights distribution toward negative values (the average weight is around zero), but still the distribution becomes markedly skewed, with long tails (see panel C in Fig. \ref{WeightsALL}).

\subsection{Strengths Distribution}
We also analyzed the strengths distribution of the network before and after learning. The strength for a given node $i$ is given by $s_i=\sum_j W_{ij}$. The distribution of each layer at time $t=0$ is still Gaussian as the  strength is just the sum of the random weights of the links connected to that node, and these are Gaussian distributed at time $t=0$. As expected from the results on the final weights, after learning the neurons displayed an overall negative strength. However, we also found elongated positive tails, especially in the layer $h1-v$, $h1-h2$ (considering links going from $h1$ to $h2$), $h2-h1$ (links going from $h2$ to $h1$), reaching also very high values (up to $s_i\approx 30,40$). The strengths of nodes in the $v-h1$ layer were all negative, indicating that nodes in the visible layers on average operated as inhibitors to nodes in the $h1$ layer. In the last hidden layer, the strength distribution displayed a high peak around $s_i\approx 10$. As we will explain later, this is a signature of the strong redundancy that has been found in the last layer's neurons: after learning, many units tend to fall into an almost identical set of features. 

\subsection{Receptive Fields} \label{RF_description}
At the end of the learning phase, hidden neurons in the network developed a variety of receptive fields that represent the set of visual features used to efficiently encode the statistical information contained in the training distribution. In particular, neurons in the first hidden layer developed receptive fields tuned to simple, localized spatial structures (such as blobs and small strokes), which were combined by neurons in the deepest layers in order to produce more complex visual features such as edge detectors and digit shapes. Some neurons, especially in the third hidden layer, learned features that were not location specific and covered the whole visual field.

\begin{figure}[t] 
 \centering
 \includegraphics[width=0.9\linewidth]{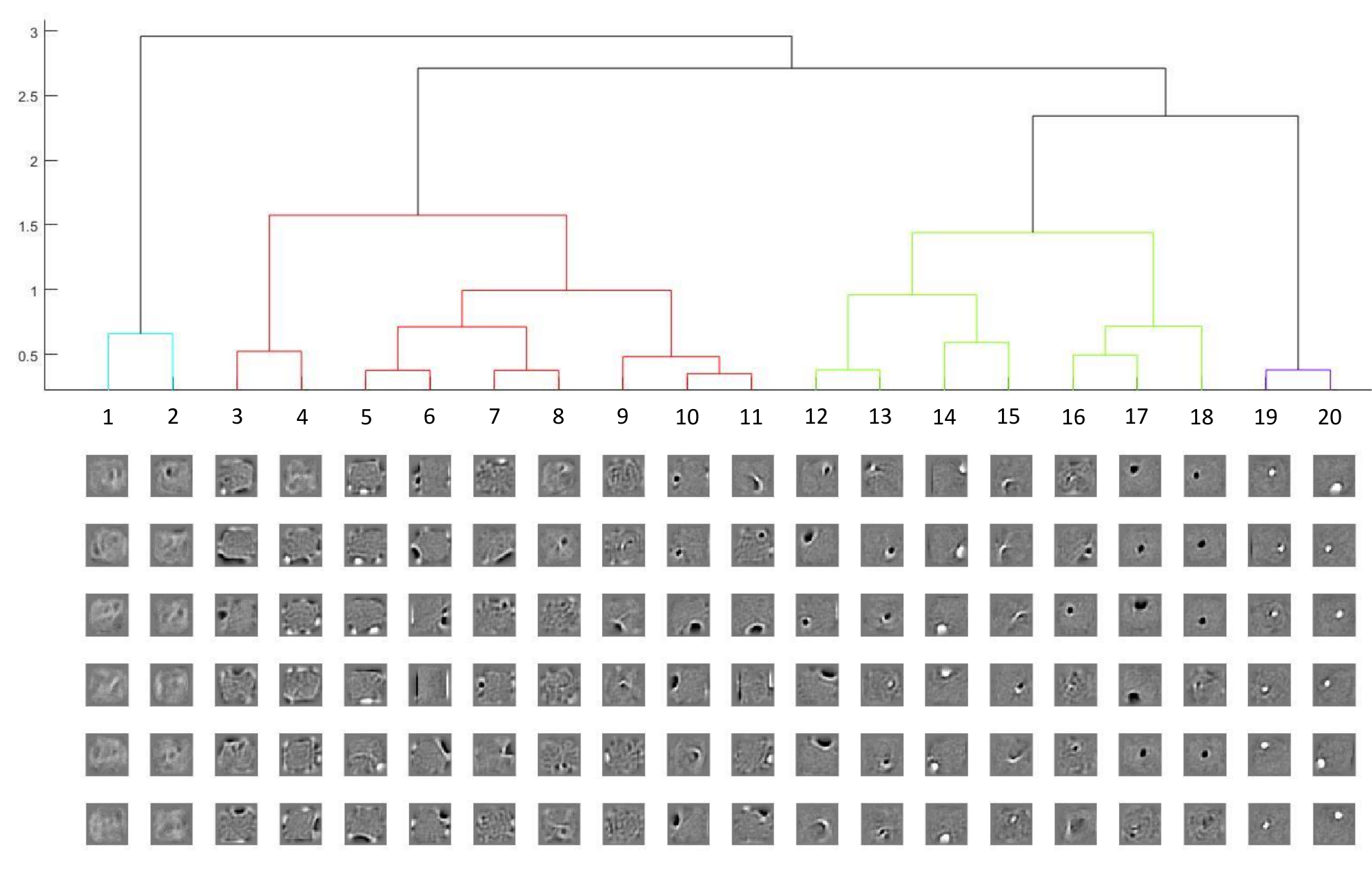}
 \caption{Hierarchical clustering of the neuronal receptive fields of the first hidden layer emerged from the MNIST handwritten dataset. The tree structure represents the distances between each of the 20 clusters, with smaller values indicating more similar types of receptive fields.}
 \label{RF_dendrogram_MNIST}
\end{figure}

\begin{figure}[h] 
 \centering
 \includegraphics[width=0.9\linewidth]{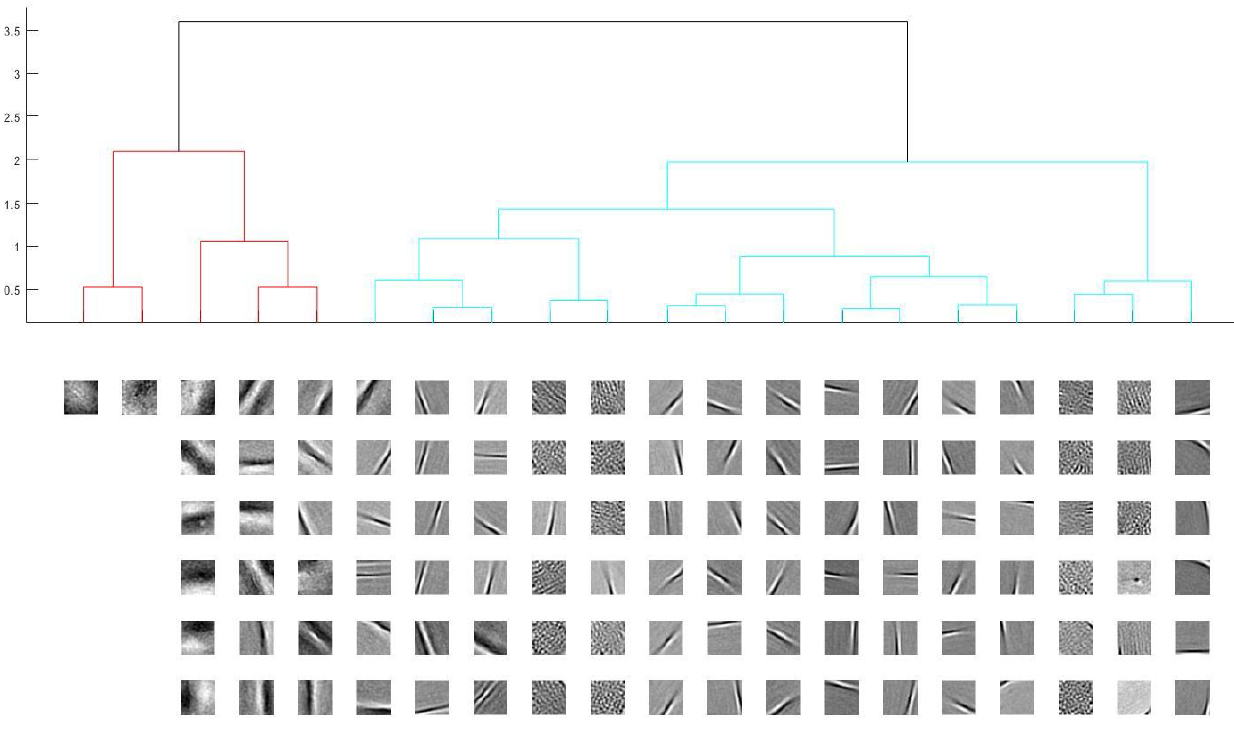}
 \caption{Hierarchical clustering of the neuronal receptive fields of the first hidden layer emerged from the natural images dataset. The tree structure represents the distances between each of the 20 clusters, with smaller values indicating more similar types of receptive fields.} \label{RF_dendrogram_Natural}
\end{figure}

After applying the clustering algorithm to the receptive fields of each hidden layer, we plotted a sample of receptive fields belonging to each cluster in order to verify that neurons encoding similar features were grouped together. As shown in Fig. \ref{RF_dendrogram_MNIST}, neurons with a similar functional role were indeed assigned to the same cluster (images in each column represent the receptive fields of the neurons belonging to the cluster identified by the numeric label). A similar result holds for the network trained on natural images (see Fig. \ref{RF_dendrogram_Natural}).

\subsection{Relation between Network Structure and Function}
\begin{figure}[t]
 \centering
 \includegraphics[width=1\linewidth]{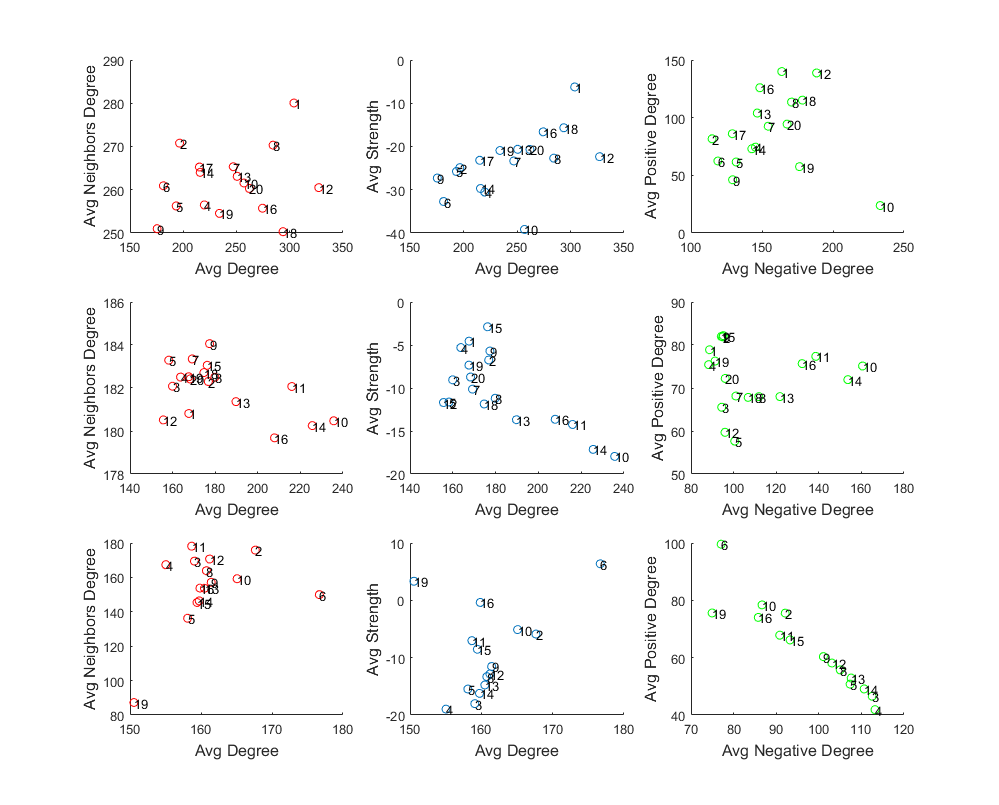}
 \caption{Structural properties of the sub-graphs $G_i$ of the deep learning system trained with the MINST dataset, for a simplification threshold of $\theta=0.1$, where each number denotes a ``functional" group of nodes defined for the similarity of the corresponding receptive fields, as described in the main text. In particular we show the relation between: \textit{(first column)} averaged nearest neighbor degrees and node degree (also known as (diss)assortativity \cite{Newman2010}), \textit{(second column)} averaged node strength and degree and \textit{(third column)} averaged positive and negative degrees. Each row corresponds to a different layer of the deep network, ordered from the first (top row) to the third (bottom row).} \label{Corr01NoOutliers}
\end{figure}

In order to unveil a potential relation between node topological properties (network structure) and node receptive fields (network function), we first considered the subnetwork formed by nodes belonging to the same functional class $\mathcal{G}_i$, for $i=1,...,G$ (with $G=19$ as described in section \ref{RF_description}). We then studied the following topological properties of the nodes for each sub-network:
\begin{itemize}
\item The average node degree $\langle k \rangle_j = \sum_{j \in \mathcal{G}_i} k_j /|G_i|$ for $j$=1..,$G$ where $|G_i|$ is the number of nodes in the subgraph $G_i$. We can also divide it into two subgroups: the average positive degree ($\langle k^+ \rangle= \sum_{j \in \mathcal{G}_i} k_j \Theta(k_j) /|G_i|$) and the averaged negative degree ($\langle k^- \rangle= \sum_{j \in \mathcal{G}_i} k_j \Theta(-k_j) /|G_i|$), where $\Theta(x)$ is the Heaviside theta function.  
\item The average nearest neighbor degree $\langle k_{nn}\rangle = \sum_{i \in \mathcal{G}_i} \sum_{j \in\,nn(i)} k_j/G$, where $nn(i)$ denotes the nearest neighbors of the node $i$. 
\item The average node strength $\langle s \rangle = \sum_{i \in \mathcal{G}_i} s_i / G$. 
\end{itemize}
We have also analyzed other properties, such as the standard deviation of the above quantities, the related coefficient of variation, the eccentricity, and other centralities measures, but they did not supply any additional relevant information to the overall picture. Figure \ref{Corr01NoOutliers} shows the results of the analysis. We note that we have removed from the analysis few subgraphs that were composed by very few nodes (less than 10), as for such clusters it was not possible to compute an average statistically meaningful behavior.

\begin{figure}[t] 
 \centering
 \includegraphics[width=1\linewidth]{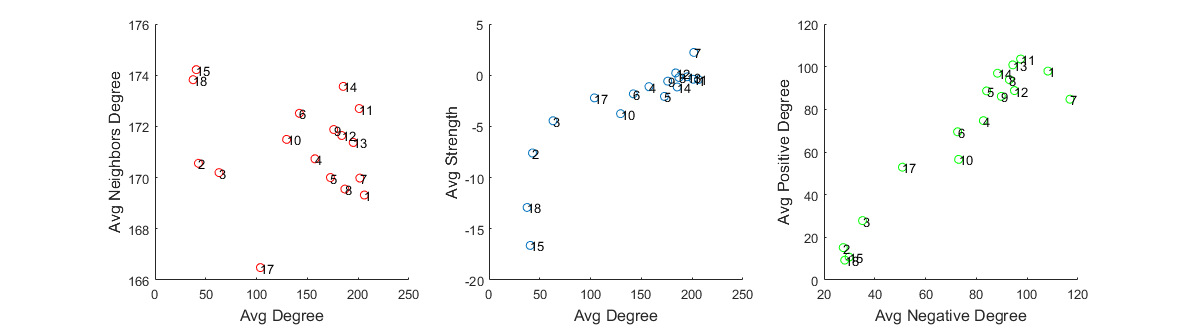}
 \caption{Structural properties of the sub-graphs $G_i$ of the deep learning system trained the natural images dataset, for a simplification threshold of $\theta=0.1$, where each number denotes a functional group of nodes defined for the similarity of the corresponding receptive fields, as the describe in the main text. In particular we show the relation between: \textit{(first column)} averaged nearest neighbor degrees and node degree, \textit{(second column)} averaged node strength and degree and \textit{(third column)} averaged positive and negative degrees.} \label{Corr01Natural}
\end{figure}

Although a clear emergent pattern is missing, some common trends have been found.
We can see that in all layers there is not a clear trend in the relation between the averaged degree and the nearest neighbor degree (leftmost column in Fig. \ref{Corr01NoOutliers}). Therefore, it seems that the deep learning system does not develop any assortativity pattern following the training process. This holds also when the system is trained using natural images (see leftmost column in Fig. \ref{Corr01Natural}).

The relationship between the average node strengths and average node degrees is informative about the relationship between topology and networks weights. In fact, if there is no correlation between the degree of vertices and the weight of edges, then the weights $w_{ij}$ are on average independent of node $i$ and $j$, and in this case it can be shown \cite{barrat2007} that the strength of a vertex is simply proportional to its degree, and thus node degree and strength provide the same information on the system. In our case, we found that the trained neural networks (both on the handwritten digits and the natural images) developed a non trivial relationship between node degrees and strengths (central column in Fig. \ref{Corr01NoOutliers} and Fig. \ref{Corr01Natural}). In particular, the first layer in the DBN displays a positive Pearson correlation ($\rho=0.56$, $p$-value = 0.019), while in the second layer we find a negative correlation ($\rho=-0.71$, $p$-value = 0.001). The network trained on natural images displays a strong, positive correlation ($\rho=0.87$, $p$-value = $<$ 0.0001).

Finally, by analyzing the relationship between the positive and negative averaged node degrees in the deep network we found that, depending on the layer, we have different results (see rightmost column in Fig. \ref{Corr01NoOutliers} and Fig. \ref{Corr01Natural}). In the first layer, with the exception of the $G_{10}$ cluster, we have a positive relationship between the two measures, quantified by a correlation of $\rho=0.59$, $p$-value = 0.016. On the other hand, in the third layer the opposite is true, and we find a clear negative correlation ($\rho=-0.93$, $p$-value $<$ 0.0001). In the second layer, no statistical correlation is detected, and there is not significant relation between the two types of degree. The relationship obtained using the neural network trained on natural images (see rightmost panel in Fig. \ref{Corr01Natural}) displays the same positive correlation found in the first layer of the deep neural network trained with the handwritten digits dataset ($\rho=0.95$, $p$-value $<$ 0.0001).

\section{Discussion and Conclusions}

In this research work we analyzed deep learning systems from a network science perspective, by investigating a variety of structural and functional properties of the emergent computational graph. Our analyses allowed to gain interesting insights about the internal functioning of these complex networks, suggesting that the proposed approach might be useful to better understand the principles governing these non-linear, self-organizing systems.

The ambitious goal of finding structural signatures of the functioning of deep learning systems turned out to be a challenging and delicate point of our analysis. First, it should be noted that in order to perform a non-trivial analysis of the topological properties of the trained networks we need to set a threshold on the connection weights. Otherwise, we would find a trivial ``all nodes connected to all nodes'' structure. Of course, setting a threshold is always a delicate operation. Our rationale has been to set a threshold that enabled to remove all the weaker links, while preserving the most relevant ones. We have performed a sensibility analysis with other threshold values ($\theta=0.01, 0.03, 0.05, 0.2$), and the main results presented here are robust with respect to the choice of different threshold values.

The emergence of inhibitory links in the network trained with the handwritten digits dataset, especially in the first and third layers, induces strong anti-correlations between neurons in the visible and top layer with those in the second one. This effect may be promoted by the type of input (i.e. the pixels distribution in handwritten digits), as this clear shift of the edge weights toward negative values disappears when the system is trained on natural images. For example, the MINST dataset contains images of white digits written on a uniform, black background: The marked contrast between elements in the image may have induced strong anti-correlations among neurons.
Indeed, we observe that many neurons settle their weights to negative values, thereby inhibiting the activity of several connected neurons. Looking at the receptive fields of these neurons, we observe that they tend to specialize in describing very localized features, thereby activating in response to very particular features in the stimulus (e.g. spots or straight edges), while they are anti-correlated with all the remaining pixels in the input image.

\begin{figure}[t]
 \centering
 \includegraphics[width=0.6\linewidth]{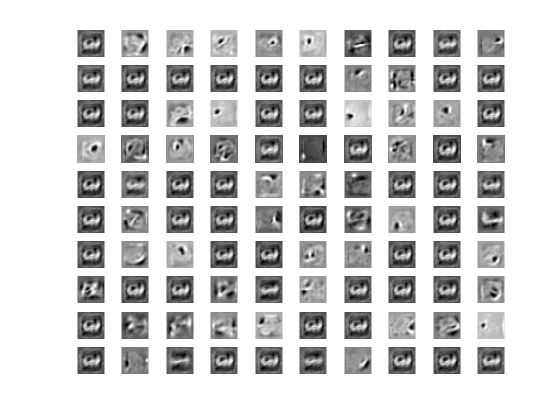}
 \caption{Some receptive fields of the third hidden layer, highlighting the presence of many dead units.} \label{lay3RF}
\end{figure}

Another relevant information can be drawn from the strength distributions of the neurons in the third layer, where a high peak of neurons with negative strength (between $s_i\in[-8,-10]$) is observed. The corresponding receptive fields are shown in Figure \ref{lay3RF}, which highlights a high level of redundancy in the neuron's function. In particular, all the nodes with a strength in the range of the peak have exactly the same receptive field, and are usually known as \emph{dead units} \cite{fritzke1995growing}. This redundancy may be due to the use of too many neurons in the third hidden layer, which might not be all necessary to improve the representational capability of the network.

A further non-trivial point is how to define the functional modules of a deep neural network. There is not an \emph{a priori} definition about the function of the neurons in a deep learning system, as the overall activity of the network nodes is very complex and cannot be classified, for example, by a binary variable (e.g. inhibitory / excitatory neuron). Here we have proposed to approximately characterize the neuron function by visualizing its receptive field. The receptive field is a complex, non trivial emerging description of the overall activity of each neuron in a given layer. Therefore, it is a highly abstract representation of the neuron function and it might not always have a clear interpretation. However, we have proposed that functionally similar neurons can be detected by clustering the corresponding receptive fields, thereby allowing to define functional sub-networks. The analysis of the topological properties and weighted architectures of these subgraphs hardly gives clear explanations on how the neural network works.

In conclusion, in this work we have proposed a network science perspective to unveil topological and functional properties of deep learning systems. Although the relation between structure and function remains only outlined, it is a first step to go \emph{beyond the black-box use} of such learning systems. In particular, our work highlights how some topological properties (i.e. emergence of inhibitory links) depend on the type of input signals, and thus might be initialized in a non-random way that is closer to the configuration observed after training. Other properties, such as system redundancy, do not depend on the input distribution, but might instead depend on the architecture of the system itself (i.e. number of neurons in each layer, or the inclusion of sparsity constraints). An interesting future perspective will be to relate these results with the recently proposed hypothesis of criticality in deep learning \cite{song2017emergence}. A related issue would be to characterize the stability of learning in deep neural networks with respect to random attacks and link failure \cite{kurant2007error,cohen2010complex,Newman2010}: how many edges can we delete before the learning system will stop working?

We believe that our investigation represents a small step toward the challenging goal of developing analytical techniques for interpreting and understanding deep learning systems \cite{samek2017explainable,lipton2016mythos}. Even though deriving analytical descriptions of such complex, self-organizing systems might seem daunting, it has been recognized as one of the most fundamental issues to be solved in the near future \cite{shneiderman2016opinion}. Indeed, these powerful AI systems are already operating in our societies, and international regulatory agencies are pressing scientists and engineers to ensure that AI systems will produce human-understandable explanations of their automated decisions \cite{goodman2016eu}.

\section*{Acknowledgments}
A.T. and S.S. was supported by the STARS grant (DEEPMATH and ReACT, respectively) from the University of Padova. A.T. and S.S. gratefully acknowledges the support of NVIDIA Corporation with the donation of a Titan Xp GPU used for this research.

\newpage



\begin{thebibliography}{10}

\bibitem{ackley1985}
David~H Ackley, Geoffrey~E Hinton, and Terrence~J Sejnowski.
\newblock A learning algorithm for boltzmann machines.
\newblock {\em Cognitive science}, 9(1):147--169, 1985.

\bibitem{agliari2015topological}
Elena Agliari, Adriano Barra, Andrea Galluzzi, Francesco Guerra, Daniele
  Tantari, and Flavia Tavani.
\newblock Topological properties of hierarchical networks.
\newblock {\em Physical Review E}, 91(6):062807, 2015.

\bibitem{Albert2002}
R{\'{e}}ka Albert and Albert-L{\'{a}}szl{\'{o}} Barab{\'{a}}si.
\newblock {Statistical mechanics of complex networks}.
\newblock {\em Reviews of Modern Physics}, 74(1):47--97, 2002.

\bibitem{Baldassi2016}
Carlo Baldassi, Christian Borgs, Jennifer~T Chayes, Alessandro Ingrosso, Carlo
  Lucibello, Luca Saglietti, and Riccardo Zecchina.
\newblock {Unreasonable effectiveness of learning neural networks: From
  accessible states and robust ensembles to basic algorithmic schemes}.
\newblock 2016.

\bibitem{baldi2014}
Pierre Baldi, Peter Sadowski, and Daniel Whiteson.
\newblock Searching for exotic particles in high-energy physics with deep
  learning.
\newblock {\em Nature Communications}, 5, 2014.

\bibitem{barrat2007}
Alain Barrat, Marc Barthelemy, and Alessandro Vespignani.
\newblock The architecture of complex weighted networks: Measurements and
  models.
\newblock In {\em Large Scale Structure And Dynamics Of Complex Networks: From
  Information Technology to Finance and Natural Science}, pages 67--92. World
  Scientific, 2007.

\bibitem{Bressler2010}
Steven~L Bressler and Vinod Menon.
\newblock {Large-scale brain networks in cognition: emerging methods and
  principles.}
\newblock {\em Trends in cognitive sciences}, 14(6):277--90, jun 2010.

\bibitem{Bullmore2009}
Ed~Bullmore and Olaf Sporns.
\newblock {Complex brain networks: graph theoretical analysis of structural and
  functional systems.}
\newblock {\em Nature reviews. Neuroscience}, 10(3):186--98, mar 2009.

\bibitem{gary2008}
Gary Chartrand and Ping Zhang.
\newblock {\em Chromatic Graph Theory}.
\newblock CRC Press, 2008.

\bibitem{cohen2010complex}
Reuven Cohen and Shlomo Havlin.
\newblock {\em Complex networks: structure, robustness and function}.
\newblock Cambridge university press, 2010.

\bibitem{Collobert2011}
Ronan Collobert, Jason Weston, Leon Bottou, Michael Karlen, Koray Kavukcuoglu,
  and Pavel Kuksa.
\newblock {Natural Language Processing (almost) from Scratch}.
\newblock {\em Journal of machine learning research}, 12:2493--2537, mar 2011.

\bibitem{fritzke1995growing}
Bernd Fritzke.
\newblock A growing neural gas network learns topologies.
\newblock In {\em Advances in neural information processing systems}, pages
  625--632, 1995.

\bibitem{Goodfellow2016}
Ian Goodfellow, Yoshua Bengio, and Aaron Courville.
\newblock {\em Deep Learning}.
\newblock MIT Press, 2016.
\newblock \url{http://www.deeplearningbook.org}.

\bibitem{goodman2016eu}
Bryce Goodman and Seth Flaxman.
\newblock European union regulations on algorithmic decision-making and a
  “right to explanation”.
\newblock In {\em ICML workshop on human interpretability in machine learning,
  New York, NY.}, 2016.

\bibitem{Guclu2015}
Umut G{\"{u}}{\c{c}}l{\"{u}} and Marcel A.~J. van Gerven.
\newblock {Deep Neural Networks Reveal a Gradient in the Complexity of Neural
  Representations across the Ventral Stream}.
\newblock {\em Journal of Neuroscience}, 35(27):10005--10014, 2015.

\bibitem{he2016deep}
Kaiming He, Xiangyu Zhang, Shaoqing Ren, and Jian Sun.
\newblock Deep residual learning for image recognition.
\newblock In {\em Proceedings of the IEEE conference on computer vision and
  pattern recognition}, pages 770--778, 2016.

\bibitem{hinton2002}
Geoffrey~E Hinton.
\newblock Training products of experts by minimizing contrastive divergence.
\newblock {\em Neural Computation}, 14(8), 2002.

\bibitem{Hinton2007a}
Geoffrey~E Hinton.
\newblock {Learning multiple layers of representation}.
\newblock {\em Trends in cognitive sciences}, 11(10):428--34, oct 2007.

\bibitem{hinton2012}
Geoffrey~E Hinton.
\newblock A practical guide to training restricted boltzmann machines.
\newblock In {\em Neural networks: Tricks of the trade}, pages 599--619.
  Springer, 2012.

\bibitem{hinton1995}
Geoffrey~E Hinton, Peter Dayan, Brendan~J Frey, and Radford~M Neal.
\newblock The wake-sleep algorithm for unsupervised neural networks.
\newblock {\em Science}, 268(5214):1158, 1995.

\bibitem{hinton2006fast}
Geoffrey~E Hinton, Simon Osindero, and Yee-Whye Teh.
\newblock A fast learning algorithm for deep belief nets.
\newblock {\em Neural computation}, 18(7):1527--1554, 2006.

\bibitem{Hopfield1982}
J~J Hopfield.
\newblock {Neural networks and physical systems with emergent collective
  computational abilities}.
\newblock {\em Proceedings of the National Academy of Sciences},
  79(April):2554--2558, 1982.

\bibitem{jordan2001}
Michael~I Jordan and Terrence~J Sejnowski.
\newblock {\em {Graphical Models: Foundations of Neural Computation}},
  volume~5.
\newblock MIT Press, Cambridge, MA, oct 2001.

\bibitem{kirkpatrick1983}
Scott Kirkpatrick, C~Daniel Gelatt, Mario~P Vecchi, et~al.
\newblock Optimization by simulated annealing.
\newblock {\em science}, 220(4598):671--680, 1983.

\bibitem{koller2009}
D~Koller and N~Friedman.
\newblock {\em {Probabilistic graphical models: principles and techniques}}.
\newblock The MIT Press, Cambridge, MA, 2009.

\bibitem{Kriegeskorte2015}
Nikolaus Kriegeskorte.
\newblock {Deep Neural Networks: A New Framework for Modeling Biological Vision
  and Brain Information Processing}.
\newblock {\em Annual Review of Vision Science}, 1(1):417--446, 2015.

\bibitem{Krizhevsky2012}
Alex Krizhevsky, Ilya Sutskever, and Geoffrey~E Hinton.
\newblock {ImageNet classification with deep convolutional neural networks}.
\newblock {\em Advances in Neural Information Processing Systems}, 24:609--616,
  2012.

\bibitem{kurant2007error}
Maciej Kurant, Patrick Thiran, and Patric Hagmann.
\newblock Error and attack tolerance of layered complex networks.
\newblock {\em Physical Review E}, 76(2):026103, 2007.

\bibitem{lecun1998}
Yann LeCun.
\newblock The mnist database of handwritten digits.
\newblock {\em http://yann. lecun. com/exdb/mnist/}, 1998.

\bibitem{LeCun2015}
Yann LeCun, Yoshua Bengio, and Geoffrey~E Hinton.
\newblock {Deep learning}.
\newblock {\em Nature}, 521(7553):436--444, may 2015.

\bibitem{Lee2008}
Honglak Lee, Chaitanya Ekanadham, and Andrew~Y Ng.
\newblock {Sparse deep belief net models for visual area V2}.
\newblock {\em Advances in Neural Information Processing Systems},
  20:873----880, 2008.

\bibitem{lin2017}
Henry~W Lin, Max Tegmark, and David Rolnick.
\newblock Why does deep and cheap learning work so well?
\newblock {\em Journal of Statistical Physics}, 168(6):1223--1247, 2017.

\bibitem{lipton2016mythos}
Zachary~C Lipton.
\newblock The mythos of model interpretability.
\newblock In {\em ICML workshop on human interpretability in machine learning,
  New York, NY.}, 2016.

\bibitem{Ma2015}
Junshui Ma, Robert~P Sheridan, Andy Liaw, George~E Dahl, and Vladimir Svetnik.
\newblock {Deep neural nets as a method for quantitative structure-activity
  relationships.}
\newblock {\em Journal of chemical information and modeling}, 55(2):263--74,
  feb 2015.

\bibitem{McClelland2010}
James~L McClelland, Matthew~M Botvinick, David~C Noelle, David~C Plaut,
  Timothy~T Rogers, Mark~S Seidenberg, and Linda~B Smith.
\newblock Letting structure emerge: connectionist and dynamical systems
  approaches to cognition.
\newblock {\em Trends in Cognitive Sciences}, 14(8):348--56, aug 2010.

\bibitem{McCulloch1943}
Warren~S. McCulloch and Walter Pitts.
\newblock {A logical calculus of the ideas immanent in nervous activity}.
\newblock {\em The Bulletin of Mathematical Biophysics}, 5(4):115--133, dec
  1943.

\bibitem{Medaglia2015}
John~D. Medaglia, Mary-Ellen Lynall, and Danielle~S Bassett.
\newblock {Cognitive Network Neuroscience}.
\newblock {\em Journal of Cognitive Neuroscience}, 27(8):1471----1491, 2015.

\bibitem{mhaskar2017}
Hrushikesh Mhaskar, Qianli Liao, and Tomaso~A Poggio.
\newblock When and why are deep networks better than shallow ones?
\newblock In {\em AAAI}, pages 2343--2349, 2017.

\bibitem{Mnih2015}
Volodymyr Mnih, Koray Kavukcuoglu, David Silver, Andrei~a. Rusu, Joel Veness,
  Marc~G. Bellemare, Alex Graves, Martin Riedmiller, Andreas~K. Fidjeland,
  Georg Ostrovski, Stig Petersen, Charles Beattie, Amir Sadik, Ioannis
  Antonoglou, Helen King, Dharshan Kumaran, Daan Wierstra, Shane Legg, and
  Demis Hassabis.
\newblock {Human-level control through deep reinforcement learning}.
\newblock {\em Nature}, 518(7540):529--533, feb 2015.

\bibitem{Mohamed2012a}
Abdelrahman Mohamed, George~E Dahl, and Geoffrey~E Hinton.
\newblock {Acoustic Modeling Using Deep Belief Networks}.
\newblock {\em IEEE Transactions on Audio, Speech, and Language Processing},
  20(1):14--22, jan 2012.

\bibitem{Newman2010}
M~Newman.
\newblock {\em {Networks: An Introduction}}.
\newblock Oxford University Press, 2010.

\bibitem{Park2013}
H.-J. Park and Karl~J Friston.
\newblock {Structural and Functional Brain Networks: From Connections to
  Cognition}.
\newblock {\em Science}, 342(6158):1238411--1238411, oct 2013.

\bibitem{raina2009}
Rajat Raina, Anand Madhavan, and Andrew~Y Ng.
\newblock Large-scale deep unsupervised learning using graphics processors.
\newblock In {\em Proceedings of the 26th annual international conference on
  machine learning}, pages 873--880. ACM, 2009.

\bibitem{rosenblatt1958}
Frank Rosenblatt.
\newblock The perceptron: A probabilistic model for information storage and
  organization in the brain.
\newblock {\em Psychological review}, 65(6):386, 1958.

\bibitem{rubner2000earth}
Yossi Rubner, Carlo Tomasi, and Leonidas~J Guibas.
\newblock The earth mover's distance as a metric for image retrieval.
\newblock {\em International journal of computer vision}, 40(2):99--121, 2000.

\bibitem{Rumelhart1986a}
David~E Rumelhart and James~L McClelland.
\newblock {\em {Parallel Distributed Processing: Explorations in the
  Microstructure of Cognition. Volume 1: Foundations}}, volume~1.
\newblock MIT Press, Cambridge, MA, 1986.

\bibitem{salakhutdinov2015learning}
Ruslan Salakhutdinov.
\newblock Learning deep generative models.
\newblock {\em Annual Review of Statistics and Its Application}, 2:361--385,
  2015.

\bibitem{samek2017explainable}
Wojciech Samek, Thomas Wiegand, and Klaus-Robert M{\"u}ller.
\newblock Explainable artificial intelligence: Understanding, visualizing and
  interpreting deep learning models.
\newblock {\em arXiv preprint arXiv:1708.08296}, 2017.

\bibitem{Schmidhuber2015}
J{\"{u}}rgen Schmidhuber.
\newblock {Deep Learning in neural networks: An overview}.
\newblock {\em Neural Networks}, 61:85--117, 2015.

\bibitem{shneiderman2016opinion}
Ben Shneiderman.
\newblock Opinion: The dangers of faulty, biased, or malicious algorithms
  requires independent oversight.
\newblock {\em Proceedings of the National Academy of Sciences},
  113(48):13538--13540, 2016.

\bibitem{Silver2016}
David Silver, Aja Huang, Chris~J. Maddison, Arthur Guez, Laurent Sifre, George
  van~den Driessche, Julian Schrittwieser, Ioannis Antonoglou, Veda
  Panneershelvam, Marc Lanctot, Sander Dieleman, Dominik Grewe, John Nham, Nal
  Kalchbrenner, Ilya Sutskever, Timothy Lillicrap, Madeleine Leach, Koray
  Kavukcuoglu, Thore Graepel, and Demis Hassabis.
\newblock {Mastering the game of Go with deep neural networks and tree search}.
\newblock {\em Nature}, 529(7587):484--489, 2016.

\bibitem{snavely2006}
Noah Snavely, Steven~M Seitz, and Richard Szeliski.
\newblock Photo tourism: exploring photo collections in 3d.
\newblock In {\em ACM transactions on graphics (TOG)}, volume~25, pages
  835--846. ACM, 2006.

\bibitem{song2017emergence}
Juyong Song, Matteo Marsili, and Junghyo Jo.
\newblock Emergence and relevance of criticality in deep learning.
\newblock {\em arXiv preprint arXiv:1710.11324}, 2017.

\bibitem{Testolin2017}
Alberto Testolin, Michele {De Filippo De Grazia}, and Marco Zorzi.
\newblock {The role of architectural and learning constraints in neural network
  models: A case study on visual space coding}.
\newblock {\em Frontiers in Computational Neuroscience}, 11(March):1--17, 2017.

\bibitem{testolin2013}
Alberto Testolin, Ivilin Stoianov, Michele De~Filippo De~Grazia, and Marco
  Zorzi.
\newblock Deep unsupervised learning on a desktop pc: a primer for cognitive
  scientists.
\newblock {\em Frontiers in psychology}, 4, 2013.

\bibitem{Testolin2017b}
Alberto Testolin, Ivilin Stoianov, and Marco Zorzi.
\newblock {Letter perception emerges from unsupervised deep learning and
  recycling of natural image features}.
\newblock {\em Nature Human Behaviour}, 1(9):657--664, sep 2017.

\bibitem{testolin2014machine}
Alberto Testolin, Marco Zanforlin, Michele De~Filippo De~Grazia, Daniele
  Munaretto, Andrea Zanella, Marco Zorzi, and Michele Zorzi.
\newblock A machine learning approach to qoe-based video admission control and
  resource allocation in wireless systems.
\newblock In {\em Ad Hoc Networking Workshop (MED-HOC-NET), 2014 13th Annual
  Mediterranean}, pages 31--38. IEEE, 2014.

\bibitem{Testolin2016a}
Alberto Testolin and Marco Zorzi.
\newblock {Probabilistic Models and Generative Neural Networks: Towards an
  Unified Framework for Modeling Normal and Impaired Neurocognitive Functions}.
\newblock {\em Frontiers in Computational Neuroscience}, 10(73), jul 2016.

\bibitem{welling2002}
Max Welling and Geoffrey~E Hinton.
\newblock A new learning algorithm for mean field boltzmann machines.
\newblock In {\em International Conference on Artificial Neural Networks},
  pages 351--357. Springer, 2002.

\bibitem{Xiong2015}
H.~Y. Xiong, B.~Alipanahi, L.~J. Lee, H.~Bretschneider, D.~Merico, R.~K.~C.
  Yuen, Y.~Hua, S.~Gueroussov, H.~S. Najafabadi, T.~R. Hughes, Q.~Morris,
  Y.~Barash, a.~R. Krainer, N.~Jojic, S.~W. Scherer, B.~J. Blencowe, and B.~J.
  Frey.
\newblock {The human splicing code reveals new insights into the genetic
  determinants of disease}.
\newblock {\em Science}, 347(6218):144--154, dec 2015.

\bibitem{Zorzi2017a}
Marco Zorzi and Alberto Testolin.
\newblock {An emergentist perspective on the origin of number sense}.
\newblock {\em Philosophical Transactions of the Royal Society B: Biological
  Sciences}, 2018.

\bibitem{zorzi2013}
Marco Zorzi, Alberto Testolin, and Ivilin~P Stoianov.
\newblock Modeling language and cognition with deep unsupervised learning: a
  tutorial overview.
\newblock {\em Frontiers in psychology}, 4, 2013.

\bibitem{Zorzi2015}
Michele Zorzi, Andrea Zanella, Alberto Testolin, Michele {De Filippo De
  Grazia}, and Marco Zorzi.
\newblock {Cognition-Based Networks: A New Perspective on Network Optimization
  Using Learning and Distributed Intelligence}.
\newblock {\em IEEE Access}, 3:1512 -- 1530, 2015.

\end{thebibliography}
\end{document}